# EasyNER: A Customizable Easy-to-Use Pipeline for Deep Learning- and Dictionary-based Named Entity Recognition from Medical and Life Science Texts

Rafsan Ahmed[1, 5], Petter Berntsson[1], Alexander Skafte[1], Salma Kazemi Rashed[1, 5], Marcus Klang[2], Adam Barvesten[1,*], Ola Olde[1,*], William Lindholm[1], Antton Lamarca Arrizabalaga[1], Pierre Nugues[2] and Sonja Aits[1,3-7 #]

[1] Cell Death, Lysosomes and Artificial Intelligence Group, Department of Experimental Medical Science, Faculty of Medicine, Lund University, Lund, Sweden

[2] Department of Computer Science, Faculty of Engineering, Lund University, Lund, Sweden

[3] Lund University Cancer Centre (LUCC)

[4] Profile area "Nature-based Future Solutions", Lund University, Lund, Sweden

[5] Profile area "Natural and Artificial Cognition", Lund University, Lund, Sweden

[6] Profile area "Proactive Ageing", Lund University, Lund, Sweden

[7] Strategic research area "BECC: Biodiversity and Ecosystem services in a Changing Climate"

* Equal contribution

[#] Corresponding author: Sonja Aits, Email: sonja.aits@med.lu.se, BMC D10, 221 84 Lund, Sweden

## Abstract

*Background*

Medical and life science research generates millions of publications, and it is a great challenge for researchers to utilize this information in full since its scale and complexity greatly surpasses human reading capabilities. Automated text mining can help extract and connect information spread across this large body of literature, but this technology is not easily accessible to life scientists.

*Methods and Results*

Here, we developed an easy-to-use end-to-end pipeline for deep learning- and dictionary-based named entity recognition (NER) of typical entities found in medical and life science research articles, including diseases, cells, chemicals, genes/proteins, species and others. The pipeline can access and process large medical research article collections (PubMed, CORD-19) or raw text and incorporates a series of deep learning models fine-tuned on the HUNER corpora collection. In addition, the pipeline can perform dictionary-based NER related to COVID-19 and other medical topics. Users can also load their own NER models and dictionaries to include additional entities. The output consists of publication-ready ranked lists and graphs of detected entities and files containing the annotated texts. In addition, we provide two accessory scripts which allow processing of files in PubTator format and rapid inspection of the results for specific entities of interest. As model use cases, the pipeline was deployed on two collections of autophagy-related abstracts from PubMed and on the CORD19 dataset, a collection of 764 398 research article abstracts related to COVID-19.

*Conclusions*

The NER pipeline we present is applicable in a variety of medical research settings and makes customizable text mining accessible to life scientists.

## Abbreviation list

NER, Named Entity Recognition; NLP, Natural Language Processing;

## Introduction

Making use of the existing medical knowledge and keeping up with the high rate of publications is a major challenge. With PubMed containing over 35 million publications [1, 2], manually reading all relevant articles has become impossible. This problem intensifies during health crises, as seen with the explosion of publications on COVID-19 topics from 2020 onwards. By summer 2022, CORD-19, a database for COVID-19-related research articles, had accumulated over 1 million entries [3]. Reviewing such large literature collections is time-consuming and costly, and not even large consortia of experts can connect all the scattered pieces of information. Therefore, there is a large need for automated text mining tools that efficiently process large scientific text collections and extract relevant information that is buried within them.

Recent advances in the field of Natural Language Processing (NLP) have led to highly capable automated text mining tools [4, 5]. Such tools can e.g. classify, group or prioritize articles, generate word clouds based on content, summarize text, or extract specific terms and information connected to them.

A key step of many text mining approaches is Named Entity Recognition (NER), the detection of relevant types of keywords [6]. This can be conducted in several ways. In the dictionary-based approach, the text is compared to long lists of keywords ("dictionaries"), e.g. a list of disease names, and full or partial matches are recorded [7]. However, this approach struggles to detect unknown terms and spelling variants. Another approach, rule-based matching, matches entities based on specific word characteristics, e.g., the "@" symbol can be used to identify email addresses. Hand-crafting such rules is often time consuming, and in many cases, there are no unique characteristics that could be used to identify all terms in an entity class. A third approach is to use deep neural networks trained on large collections of texts in which entities have been labelled by experts (so-called gold-standard corpora). Deep neural networks make use of the context of each word or multi-word term to decide whether it represents an entity of interest. This approach is more forgiving for unknown terms and spelling variants [8-11]. Taking the sentence context into account also makes it easier to reliably find entities. For example, in the sentence "We measured lamp expression in the cytosol." the context makes "lamp" identifiable as a protein name and not an illumination device.

Deep learning NLP models typically have millions or even billions of trainable parameters and typically use a specific architecture called transformers [4, 12-15]. Such deep neural networks are typically not trained from scratch for a specific task, as this would require extremely large annotated corpora. Instead, networks pre-trained on very large unlabeled text collections (so-called language models) are only fine-tuned for the task of interest [16], which is referred to as transfer learning. Several language models for medical English are publicly available, with many based on the BERT architecture, e.g., BioBERT [17], Clinical BERT [18], BlueBERT [19] and PubMedBERT [20]. After fine-tuning these models for NER on annotated corpora, they detect entities such as diseases or chemicals remarkably well when evaluating them on a text collection resembling the training corpus. However, generalization to texts that do not match the training data remains a problem [21]. Furthermore, these models, even when embedded in mature NLP frameworks such as spaCy [22], Flair [23] or the Hugging Face Transformers library [24], remain usable mostly for NLP specialists or others with significant programming expertise and not for the medical researchers who need continuous access to text mining technology. Several research tools, such as the STRING protein-protein interaction database [25], EuropePMC literature database [26], or the PubTator3.0 [27] and BERN2 tools [28], present information extracted by text mining for medical researchers. However, with these tools users have little control over the text mining process. There is therefore a need for end-user-oriented text mining tools that are customizable, accessible for medical researchers and applicable across different medical research domains.

Here, we present an end-to-end pipeline for NER with integrated BioBERT models [17] fine-tuned on the large HUNER corpus collection [29]. This enables detection of terms for cells, chemicals, diseases, genes/proteins and species. The pipeline can also perform dictionary-based NER, and three COVID-19-related dictionaries, previously developed by our group [30, 31], are included. Users have full control over the input texts and can also load their own NER models or dictionaries. The pipeline outputs a ranked list of identified entities and a graph of the most frequent entities which are easy to comprehend for life scientists as well as structured annotation files for downstream analysis. Separate scripts for processing of PubTator files as input and for rapid inspection of the results for a specific entity of interest are also provided. We demonstrated the use of the pipeline in two model cases, information extraction from autophagy-related abstracts in PubMed and from the CORD-19 database.

## Material and Methods

### Computing and data storage resources

EasyNER was developed using Python version 3.9 and pytorch version 1.13 with GPU support. We recommend this and an NVIDIA GPU of series 20XX for optimal performance. The pipeline can also be run on multiple functional CPUs (threads) in parallel without using a GPU, but the runtime may slow down. EasyNER is compatible with Windows, Linux and Mac operating systems.

Runtime experiments were performed on an ASUS TUF gaming laptop A15 (FA507NV) with an NVIDIA GeForce RTX 4060 GPU which has 8 GB graphical memory. For other computing and data storage we used the Alvis HPC cluster (Chalmers University Sweden), Berzelius HPC cluster (National Supercomputer Center Linköping University), LUNARC HPC cluster (Lund University) and a variety of laptops.

### Data

*Annotated gold-standard corpora*

For model fine-tuning and evaluation, we used the HUNER corpora collection [29, 32], which contains sub-corpora with annotations for several entities relevant for medical research: cells, chemicals, diseases, genes/proteins and species (Supplemental file 1). These 5 sub-corpora were created by combining several corpora for each entity. The HUNER collection with gold-standard IOB2/CoNLL2002 [33] NER and part-of-speech annotations, was obtained using HunFlair[1] [32], with a modification made to the corpus collection code to download the OSIRIS corpus[2] [34]. This was necessary to overcome an error in the code. The HunFlair version of HUNER does not include the BioSemantics corpus that was present in the original HUNER collection. Each of the 5 HUNER sub-corpora is pre-split into training, development and test sets [32]. For our model training the part-of-speech tags were removed.

Models were also fine-tuned and evaluated on the BC5CDR_Disease corpus in IOB2 format that had been used in the BioBERT study [17, 35]. The dataset is pre-split into training ("train"), development ("dev") and test set ("test") and was extracted from the larger HUNER corpora collection.

The Lund-COVID-19 corpus contains 10 SARS-CoV2-related abstracts from the CORD-19 dataset with IOB2 NER annotations [30, 31]. The "protein" class in this corpus contains both gene and protein annotations and corresponds to the "gene" class in the HUNER corpus, which also has annotations for both entity types. We merged some of the original entity classes to obtain annotations corresponding to HUNER entities "species" (i.e. merge of Species_human, Species_other, Virus_family, Virus_other, Virus_SARS-CoV-2) and "disease" (i.e. merge of Disease_COVID_19 and Disease_other). The annotation classes "chemicals" and

[1] Retrieved from https://github.com/hu-ner/huner/tree/master/ner_scripts on Nov 4th, 2021
[2] Retrieved from https://github.com/Rostlab/nala/tree/develop/resources/corpora/osiris

"cells" were removed as there were too few entities in these classes for evaluation. This modified dataset is called the "Simplified Lund COVID-19 corpus" (Supplemental file 2)[3].

The CRAFT (Version 4.0.0) corpus contains 97 annotated articles [36]. The corpus was downloaded and converted to PubAnnotation format[4], converted from PubAnnotation format to IOB2 format with a custom script[5] which tokenized the text using the ScispaCy tokenizer (version 0.5.1, model en_core_sci_sm) [37] and then processed with the BioBERT preprocessing script[6] [17]. The max sequence length set for the BioBERT preprocessing script was kept at the default value of 192, which splits sentences larger than this length into two. "Chemical Entities of Biological Interest (CHEBI)" was used as "chemical" class, "NCBI Taxonomy (NCBITaxon)" as "species" class and "Protein Ontology (PR)" as "gene/protein" class.

The MedMentions corpus[7] contains 4392 annotated full-text articles in PubTator format randomly chosen among those released on PubMed in 2016 [38]. For benchmarking, the entity classes were remapped to the classes predicted by the EasyNER BioBERT models using a custom script[8] and mapping table (Supplemental file 3). This version of the corpus is referred to as Simplified MedMentions corpus.

The tmVar 3.0 corpus [39, 40] contains 500 annotated abstracts. The Bio-ID corpus [41] contains annotated figure panel captions from 570 articles. For both tmVar 3.0 and Bio-ID corpus we used the PubTator gold standard files released with Hunflair2[9].

The BioRED corpus[10] contains a total of 600 annotated abstracts in PubTator format, containing annotations for gene/protein, chemical, variant, disease, species and cell line entities and their relations. We used only the test set of the BioRED corpus containing 100 annotated abstracts for benchmarking.

All corpora have retained letter casing (capitalization).

---

[3] The Simplified Lund COVID-19 corpus with disease, protein and species entities is available at https://github.com/Aitslab/EasyNER/blob/main/data/Simplified%20Lund%20COVID19%20corpus.zip.

[4] Retrieved from https://github.com/UCDenver-ccp/CRAFT/releases/tag/v4.0.0 on March 24, 2023, and converted to PubAnnotation format following the instructions: https://github.com/UCDenver-ccp/CRAFT/wiki/Alternative-annotation-file-formats

[5] https://github.com/Aitslab/EasyNER/blob/main/supplementary/experiment_scripts/CRAFT_preprocessing_spacy.py

[6] Retrieved from https://github.com/dmis-lab/biobert-pytorch/blob/master/named-entity-recognition/preprocess.sh on June 8, 2021.

[7] Retrieved from https://github.com/chanzuckerberg/MedMentions on Aug 13, 2024

[8] https://github.com/Aitslab/EasyNER/blob/main/supplementary/experiment_scripts/preprocess_pubtatorformat.py

[9] Retrieved from https://github.com/hu-ner/hunflair2-experiments/tree/main/annotations/goldstandard on Aug 3, 2024

[10] Retrieved from https://ftp.ncbi.nlm.nih.gov/pub/lu/BioRED on July 15th, 2024

*Autophagy-related abstract collections*

As test cases for the pipeline, we created two collections of autophagy-related abstracts from PubMed. The first dataset, Lund Autophagy-1 (supplemental file 4), was obtained by searching PubMed with the search term “mTOR AND TSC1” on May 24, 2022. Mammalian target of rapamycin (mTOR) and Tuberous Sclerosis 1 (TSC1) are key regulators of autophagy. The second dataset, Lund Autophagy-2 (supplemental file 5) was obtained by searching PubMed on Dec 13, 2022 with the search terms “autophagy AND cancer” restricting the date to between 2020 and 2023. Both search results were exported from PubMed as individual text files containing a list of PubMed IDs and abstracts downloaded using the NER pipeline described below.

*CORD-19*

As second test case for the pipeline, we used CORD-19, a collection of coronavirus-related articles published until June 2, 2022 to aid pandemic efforts [3]. We used the final version of its metadata file published June 2, 2022[11] which holds information on 1 056 660 coronavirus-related articles including their abstracts. The CORD-19 dataset contains duplicate entries in respect to abstracts/titles and other metadata as well as entries without abstracts, both of which are removed by the NER pipeline. This yielded 764 398 unique abstracts (with title) from which entities were extracted.

**Exploratory Data Analysis**

An initial exploratory data analysis was performed for the HUNER corpora. The size of the corpus was assessed by counting the number of lines, since each line contains one token and its IOB2 tag. The number of entities was assessed by counting the number of B tags (the tag indicating the beginning of an entity) (Script in supplemental file 3).

To assess similarity between the training, development and test sets (e.g. HUNER_chemical training set vs HUNER_chemical development set), word and bi-gram frequency distribution was visualized in interactive scatter plots with a custom script (comparecorpora.py) using the scattertext tool (version 0.1.10, script in supplemental file 3) [42].

**Fine-tuning of BioBERT models**

We used the PyTorch version of the BioBERT base and large cased v. 1.1 models [17] and fine-tuned them on the combined training and development sets of the five HUNER sub-corpora, resulting in models trained to recognize a single entity. We re-used the official BioBERT training scripts[12], which perfom WordPiece tokenization. In this process, each sub-word tokens inherits the label of the original word. The default hyperparameters were used for

[11] Retrieved from https://ai2-semanticscholar-cord-19.s3-us-west-2.amazonaws.com/historical_releases.html.
[12] Retrieved from https://github.com/dmis-lab/biobert/blob/master/run_ner.py on October 29, 2021.

fine-tuning but for some models, a warmup ratio of 0.1 [43] was introduced to reduce volatility and early overfitting during training. We also implemented early stopping with a patience of 50. The maximum sequence length was set to 192. Models designated "_v1" were fine-tuned on the combined HUNER train and dev set, similarly to the HunFlair authors [32], with early stopping based on the F1 score of the test set. Models designated "_v2" were fine-tuned on the train set only, with early stopping based on the F1 score of the dev set. x

We also fine-tuned a BioBERT base cased v. 1.1 model on the BC5CDR_disease corpus train set in the same manner (including early stopping) using the same hyperparameters.

All models have been released on the HuggingFace repository ( https://huggingface.co/aitslab) with the following DOIs:

biobert_huner_cell_v1: https://doi.org/10.57967/hf/2030

biobert_huner_chemical_v1: https://doi.org/10.57967/hf/2033

biobert_huner_disease_v1: https://doi.org/10.57967/hf/2034

biobert_huner_gene_v1: https://doi.org/10.57967/hf/2031

biobert_huner_species_v1: https://doi.org/10.57967/hf/2032

biobert_bc5cdr_disease_v1: https://doi.org/10.57967/hf/3981

biobert_huner_cell_v2: https://doi.org/10.57967/hf/3789

biobert_huner_chemical_v2: https://doi.org/10.57967/hf/3786

biobert_huner_disease_v2: https://doi.org/10.57967/hf/3790

biobert_huner_gene_v2: https://doi.org/10.57967/hf/3785

biobert_huner_species_v2: https://doi.org/10.57967/hf/3788

biobert_bc5cdr_disease_v2: https://doi.org/10.57967/hf/3780

**Token-level model evaluation and benchmarking**

The fine-tuned BioBERT models were first evaluated on token-level using the corresponding HUNER test sets with the BioBERT evaluation script[13] [17] with the maximum sequence set to 192. This script in turn relies on the seqeval evaluation script in default mode [44] which is designed to mimic the results from the conlleval Perl script. In this evaluation, the predictions in IOB2 format were evaluated by comparing the B, I and O tags with the annotated "true" values. Next, the models were evaluated in the same way on fully independent IOB2-formatted datasets, the Simplified Lund COVID-19 corpus, CRAFT corpus and the BC5CDR_disease corpus test set (described above).

[13] Retrieved from https://github.com/dmis-lab/biobert-pytorch/blob/master/named-entity-recognition/run_ner.py on January 22, 2022.

For comparison we also evaluated ScispaCy [37] and HunFlair [32]. ScispaCy contains 4 multi-class NER models. We used the models fine-tuned on the BioNLP13CG (scispaCy en_ner_bionlp13cg_md, recognizes many NER classes), JNLPBA (scispaCy en_ner_jnlpba_md, recognizes cell lines, cell types, DNAs, RNAs, proteins) and CRAFT corpora (scispaCy en_ner_craft_md, recognizes cell types, chemicals, proteins, genes) [14]. HunFlair contains flair-based single-class NER models fine-tuned on the different HUNER sub-corpora.

A detailed step-by-step description of the evaluation procedure can be found in the tutorial section of the EasyNER GitHub page[15].

## Pipeline structure

An end-to-end pipeline (Figure 1) was designed to automatically access and process medical texts for NER. The pipeline includes the BioBERT models fine-tuned on the HUNER corpora and COVID-19-related dictionaries but can also load user-provided BioBERT/BERT-like models or dictionaries. The pipeline is built in modules that can also be run individually. Desired settings such as model parameters and input/output paths are defined in a config file that can be re-used and shared to ensure reproducibility. The config file also contains an option to note the runtime for each of the modules in the pipeline.

The pipeline, supporting scripts and full documentation, including installation and usage instructions, as well as tutorials for reproducing the work in this article, are provided in the EasyNER repository on GitHub[16] and CodeOcean[17].

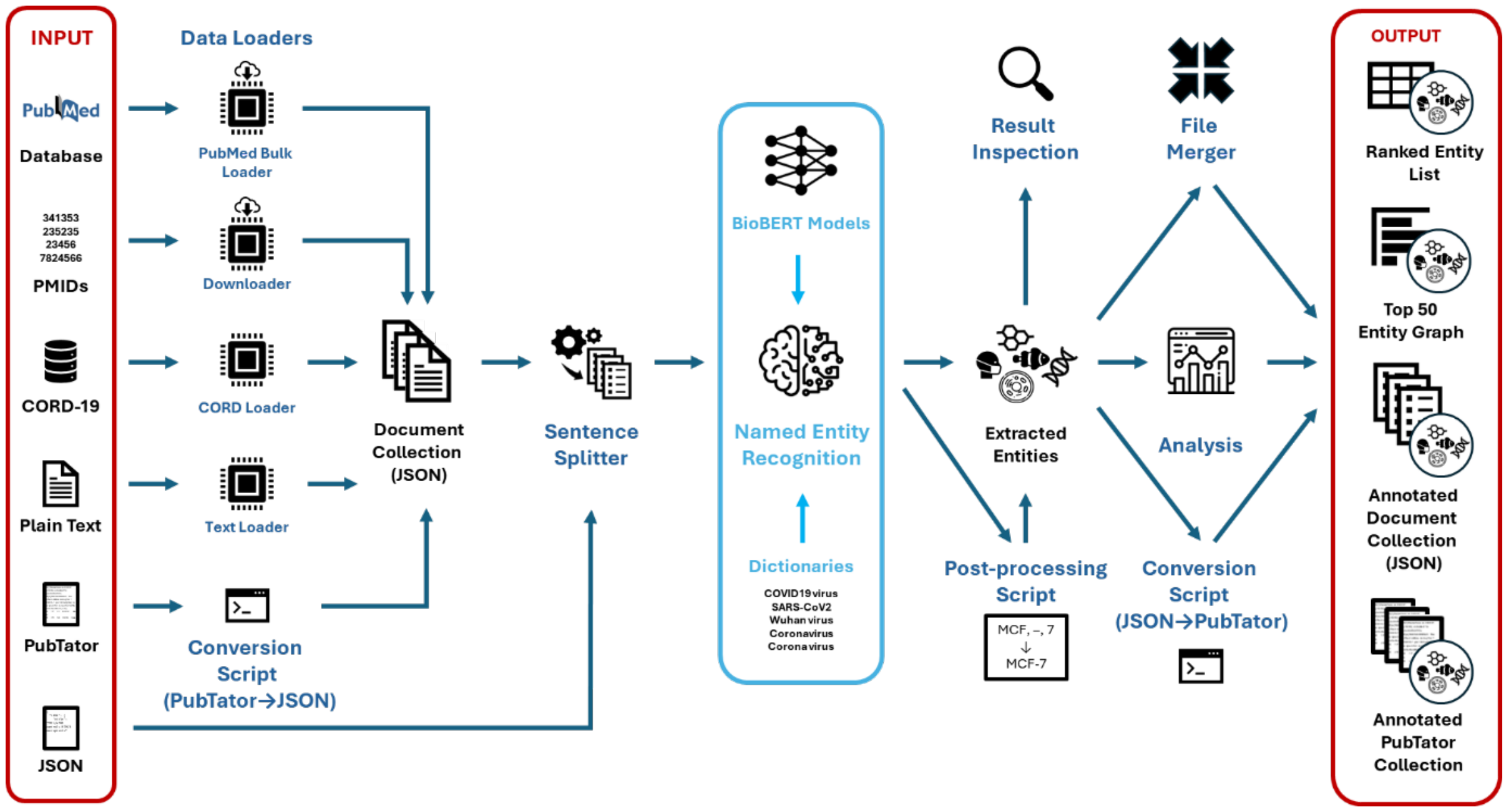

Figure 1. Overview of the EasyNER medical NER pipeline. EasyNER is built up of different modules which can be run together in sequence or individually for maximum flexibility. A variety of input formats can be processed using a set of Data

[14] Retrieved from https://github.com/allenai/scispaCy on March 24, 2023.
[15] https://github.com/Aitslab/EasyNER/blob/main/tutorials/Tutorial-evaluation_tokenlevel.md
[16] https://github.com/Aitslab/EasyNER/
[17] https://codeocean.com/capsule/1537434/

Loaders. When working with PubMed abstracts as input the user can either provide a list of PMIDs, which EasyNER accesses automatically (Downloader), or automatically download the entire PubMed database or a subset (PubMed Bulk Loader). Each document is split into sentences (Sentence Splitter module) and subsequently annotated by the Named Entity Recognition (NER) module which can use the in-built BioBERT models fine-tuned on the HUNER sub-corpora or dictionaries to recognize key life science entities (chemicals, genes/proteins, diseases, species, cells and COVID-19-related terms). Users can also load their own BioBERT-like model or dictionary to expand the NER repertoire. In addition to the files with entity annotations generated by the NER module, a list and graph of the entities ranked by count can be produced with the Analysis module. Additional optional modules can be used to combine annotated files from separate pipeline runs (File merger module), inspect the results for individual entities (Result inspection module), merge fragmented entities (Post-processing script), convert the annotated files to PubTator format (Conversion Script (JSON→PubTator)) or benchmark token-level performance (Metrics module, not shown in image).

*Data Loader module*

The pipeline has data loaders for four input types: a list of PubMed IDs, PubMed database bulk files, the CORD-19 metadata file [3], or a text file. The config file includes an "ignore" section in the beginning where the user indicates the input type (see figure 1). The user can only choose one input type per run and also needs to provide input and output file paths in the config file.

When using the PubMed ID list option, a single .txt file with one ID per line needs to be provided. Such a text file can be exported from PubMed after any search. This option runs the "Downloader" script, which downloads the abstracts and associated metadata using the e-utils PubMed API[18] [2] and parses them using PubMed Parser (version 0.3.1)[19] [45]. The raw abstracts are then merged into a document collection.

When using the PubMed database bulk file option, the pipeline will download files from the 2023 PubMed annual baseline[20], which contains all records published by December 2023, and convert them to json format. By default, the entire baseline content, >35 million publications, will be downloaded but the user can also specify the baseline file number in the config file. In addition, PubMed daily update files can be downloaded and processed in the same manner.

When using the CORD-19 option, the CORD-19 metadata file [3] needs to be provided as input. This is a csv file that contains rows of titles, abstracts and additional information for each record in the CORD-19 dataset. This option runs the "CORD loader" script on the input file which removes duplicate abstracts and entries without abstracts and then extracts titles and abstracts for the remaining 764 398 unique records from the metadata file. Alternatively, it is also possible to provide an additional .txt file with a list of CORD uIDs to the CORD loader script (one ID per line) to extract only a subset of those specific abstracts. In this case, the user needs to provide the path to the subset file and provide the argument, "subset": "true" to the config file.

Plain text documents in .txt format can be processed using the Text Loader option. Word documents and other similar files need to be converted to .txt before using them with EasyNER.

[18] https://github.com/biocommons/eutils
[19] https://github.com/titipata/pubmed_parser
[20] https://ftp.ncbi.nlm.nih.gov/pubmed

As output, all of these data loaders generate a document collection in JSON format containing PubMed IDs (or dummy ID for the Text Loader option), titles and abstracts in the user provided output path. The scripts are written in such a way that if there are no PubMed IDs, no abstracts or no text available for an article, the scripts ignore the article and move onto the next one (if available). In case of duplicate entries within the input, the Downloader and CORD Loader scripts keep the last sample of the article.

When using the pipeline with PubTator files the PubTator-to-JSON conversion script[21] is used instead of the Data Loader module to produce a EasyNER document collection JSON file that can then be processed by the Sentence Splitter module.

*Sentence Splitter module*

Before feeding the JSON file with the collected text into the NER module, the text needs to be split into single sentences. The document collection file obtained from the Data Loader module is used as input for the Sentence Splitter module. The user can choose between the faster NLTK (version 3.7) [46] sentence tokenizer or the more accurate spaCy [47] sentence tokenizer with one of the two default spaCy language models, "en_core_web_sm" or "en_core_web_trf" (version 3.3.0 for both). The NLTK and spaCy tokenizers are more suitable than a simple regex sentence splitter because medical research articles often contain mid-sentence punctuations that can be easily mistaken as end of sentence by a simple regex splitter. Note that the two spaCy models produced identical splits when tested on 3000 abstracts in which they produced over 19000 sentences (data not shown) whereas NLTK produced 3% more sentences.

The Sentence Splitter processes articles in the document collection in batches. The batch size can be specified by the user in the config file. For the smaller autophagy datasets we used a size of 100 and for the much larger CORD19 dataset we used a batch size of 1000. The sentence splitter is run parallelly through the batches using python multiprocessing library.

The output of the Sentence Splitter module is a folder that contains a collection of JSON files, which each contain one batch of texts split into individual sentences. The input and output paths, batch size, filename prefix, tokenizer and model names are all to be provided in the "splitter" section of the config file.

*NER module*

In the NER module the selected NER tagger generates entity predictions on the sentences produced by the Sentence Splitter module. There are two options for this module, NER with BioBERT/BERT models or dictionary-based NER.

---

[21] https://github.com/Aitslab/EasyNER/blob/main/supplementary/experiment_scripts/convert_hunflair2_pubtator_to_json.py

For the BioBERT/BERT option, the user can choose from an integrated collection of BioBERT models, which we fine-tuned to recognize cells, chemicals, diseases, genes/proteins or species, or load their own BioBERT or BERT-like PyTorch models, by specifying the path to the model folder and the model name in the config file. For custom models, they should be in BERT model format (model folder with PyTorch binary model file, vocab file and tokenizer). This module uses the HuggingFace Transformers library (version 4.20.1). In the BioBERT/BERT NER option, the sentences are tokenized with the BioBERT tokenizer [17], which first converts the sentences into tokens understandable by the model. BioBERT uses a WordPiece tokenizer that breaks down words into sub-words present in its vocabulary [48], to handle out-of-vocabulary words. The predictions are generated by transforming the collection of sentences into a HuggingFace dataset object and subsequently applying the model to the entire dataset using mapping. After predictions, the labels of word pieces are automatically consolidated into word-level predictions using the "max" aggregation strategy in the HuggingFace Transformers Pipeline module.

The dictionary-based NER option makes use of the spaCy Phrasematcher [47]. The user can choose between three SARS-CoV2-related dictionaries, which are downloaded in the default installation of the pipeline: a dictionary containing synonyms for "COVID-19", a dictionary containing synonyms for "SARS-CoV2" and a dictionary containing SARS-CoV2 variant names (Lund COVID-19 dictionaries, version 2, from [30, 31]). Alternatively, the user can provide their own list of terms that are to be matched in a .txt file (one term per line). In the dictionary-based NER option, the sentences are tokenized with one of the default spaCy models ("en_core_web_sm" or "en_core_web_trf").

Like the Sentence splitter, the NER module runs with user-defined batches of articles – the same as the sentence splitter. For the NER module, the user does not need to specifically provide the batch size, as sentence splitter module already splits the document collection in batches. Each batch is queued to run in parallel for the predictions.

The output of the NER module is a collection of JSON files (one per batch) containing the original texts from the original documents, split into sentences, and the predicted entity annotations (referred to as "annotated JSON document collection files" below). By default, titles are excluded from the prediction. Capitalization of the detected entities is removed at this step.

*Post-processing module*

Entities containing a hyphen or brackets (regular, square or curly) can be incorrectly fragmented in the NER process. The free-standing post-processing script[22], which is not incorporated in the pipeline, processes the annotated JSON document collection files generated by the NER module and merges the fragments. The output is a new set of annotated JSON document collection files in which the fragmented entities are merged and the other entities have remained unchanged.

[22] https://github.com/Aitslab/EasyNER/blob/main/supplementary/experiment_scripts/postprocess_separator_merging.py

*File Merger module*

When several entity types are to be annotated, or BioBERT/BERT and dictionary-based annotation are to be combined for the same entity, the NER module needs to be run repeatedly, one model/dictionary at a time. The annotated JSON document collection files from these separate runs can then be merged using the optional File Merger module if they contain the same document collection (i.e. were produced with the same batch size setting). The output consists of a new set of annotated JSON document collection files containing all entities from the input files and a file indicating the overlap.

*Analysis module*

The Analysis module processes annotated JSON document collection files containing a single entity class (but not merged files with more than one entity class) and quantifies the detected entities. The output consists of a ranked entity list in tsv format (which can be opened in Excel or similar spread sheet programs) and a publication-ready bar graph of the most frequent entities.

*Result inspection module*

The optional free-standing Result Inspection script[23] filters the generated JSON files with annotations for a single entity of interest. The output consists of a new annotated JSON file containing only the sentences with this entity for rapid inspection.

*Metrics module*

The metrics module can be used to evaluate the performance of NER models/dictionaries on token level, similarly to the BioBERT evaluation script described below. It calculates precision, recall and F1 scores by comparing an IOB2-formatted file with predictions with an IOB2-formatted file with the true annotations (ground truth). Note that this was not used for the token-level evaluations in this article.

[23] https://github.com/Aitslab/EasyNER/blob/main/scripts/search.py

*JSON-to-PubTator conversion script*

A free-standing JSON-to-PubTator Conversion Script is included to convert annotated EasyNER JSON output files to PubTator format[24].

**Benchmarking**

To benchmark EasyNER, we compared entity-level prediction to that of several other publicly available BioNLP tools: ScispaCy [37], HunFlair2 [49], PubTator Central/PubTator3 [27, 50], BENT [51] and BERN2 [28].

Hunflair2 is an updated version of HunFlair [32] which performs multi-class NER for the classes cell line, chemical, disease, gene and species. HunFlair2 has been trained on the BioRED corpus for all five entity classes, NLM Gene and GNormPlus for genes, Linneaus and S800 for species, NLM Chem and SCAI Chemical for chemicals and NCBI Disease and SCAI disease for disease predictions [49].

PubTator Central and its updated version PubTator3 are web-based tools[25] that provide access to pre-annotated PubMed and PubMed Central documents. In addition, the pre-annotated files can be downloaded via ftp or an API. The API can also process user-defined input texts in BioC, PubTator or JSON format. PubTator Central relies on GNormPlus for the annotation of genes/proteins, a re-trained tmVar 2.0 (using both abstracts and full text) for genetic variants, SR4GN for species, the original TaggerOne models for diseases and cell lines and a retrained TaggerOne model for chemicals (trained on the BC5CDR corpus and the CHEMDNER corpus) [50]. PubTator3 relies on AIONER which was trained on a combination of the NLM-Gene, NLM-Chem, NCBI-Disease, BC5CDR, tmVar (Version 3), Species-800, BioID and BioRED corpora to recognize genes/proteins, chemicals, diseases, species, genetic variants, and cell lines [27]. The PubTator Central predictions published by the HunFlair2 authors were produced with the API [49].

BENT[26] is a Python package for biomedical named entity recognition and linking for the Linux operating system. It uses 10 PubMedBERT-based NER models[27], each fine-tuned on multiple corpora for a single entity class, which recognize diseases, chemicals, genes/proteins, species, cell types, cell lines, biological processes, anatomical entities, cell components and DNA/protein variants [51].

BERN2 is an updated version of BERN which detects 9 entity classes. It has a web demo and can be used as an API. In principle, it should also be possible to install it locally, but our attempts to do this failed due to unsolvable errors[28]. The BERN2 API and web demo can process plain text and PMIDs. In the latter case, pre-computed annotations are returned from its database if available, making the prediction faster. BERN2 was trained on the BC2GM corpus for gene/protein, NCBI-disease for disease, BC4CHEMD for drug/chemical, Linnaeus for species and JNLPBA for cell line, cell type, DNA and RNA predictions [28].

[24] https://github.com/Aitslab/EasyNER/blob/main/supplementary/experiment_scripts/convert_easyner_output_json_to_pubtator.py

[25] https://www.ncbi.nlm.nih.gov/CBBresearch/Lu/Demo/PubTatorCentral/

[26] https://BENT.readthedocs.io

[27] https://huggingface.co/pruas

SciSpacy is described in the token-level evaluation section.

*Entity-level NER evaluation*

Entity-level performance was evaluated on several corpora: Simplified MedMentions (described above) [38], tmVar 3.0 [39, 40], Bio-ID [41] and BioRED (test set only) [52]. Details on all corpora can be found in the "Data" section.

For benchmarking, we re-evaluated the HunfFlair2, ScispaCy, PubTator Central, BENT and BERN2 predictions in PubTator format, which had been published in the HunFlair2 repository[29]. As the BERN2 annotation files lacked the abstract texts these were added from the files in the "raw" subfolder to keep the formatting consistent for the evaluation script. without abstracts were available[30]. In addition, we made new predictions for HunFlair2 (referred to as "HunFlair2 rerun"). PubTator3 was excluded because it had been trained on most of the evaluation corpora, which would have biased the results.

To obtain EasyNER predictions, the corpora files in PubTator format were converted to single JSON files with the EasyNER PubTator-to-JSON conversion script. Predictions were then obtained by running the EasyNER pipeline repeatedly with the five different BioBERT_HUNER v1 models, followed by the post-processing module. After the EasyNER runs, the EasyNER output JSON file was converted back to PubTator format with the EasyNER JSON-to-PubTator conversion script.

A custom evaluation script incorporating large parts of the HunFlair2 evaluation script[31], was used to compare predictions to gold standard annotations and calculate entity-level false-positives and -negatives, single class precision, recall and F1 score. The script also harmonized the names of the annotated classes (e.g. "organism" was renamed to "species" and cell type and cell line annotations were merged into the class "cell"). Before running the evaluation script, we added a dummy identifier for entities lacking one ("-1") in the PubTator file using a preprocessing script[32] as entities without identifier were not loaded by the data loading function used in the evaluation script.

[28] Described in this GitHub issue: https://github.com/dmis-lab/BERN2/issues/70

[29] https://github.com/hu-ner/hunflair2-experiments/tree/main/annotations

[30] https://github.com/Aitslab/EasyNER/blob/main/supplementary/experiment_scripts/preprocess_BERN2_into_evaluation_ready_format.ipynb

[31] https://github.com/Aitslab/EasyNER/blob/main/supplementary/experiment_scripts/evaluate_ner_pubtatorformat.py

A detailed step-by-step description of the evaluation procedure can be found in the tutorial section of the EasyNER GitHub page[33].

### User experience evaluation

We performed a qualitative evaluation of the usability of EasyNER for life scientists, examining ease of setup and use and features relevant to routine use in a research context.

## Results

### Exploration of the HUNER corpora

As models fine-tuned on a single gold-standard corpus typically generalize poorly when applied to texts of a different type we chose to train the models on the diverse and large HUNER corpora collection instead. Rather than being a single corpus, HUNER combines several gold-standard corpora harmonized to IOB2 format with one token per line. The HUNER collection consists of five sub-corpora, each annotated for a single entity, namely cells (comprising generic cell terms and cell line names), genes/proteins, diseases, species and chemicals (including therapeutic drugs). These entity classes are widely applicable in medical research.

We first explored the composition of the HUNER sub-corpora. Size of the sub-corpora and number of annotated entities differed significantly (Supplemental file 1). The training set of the HUNER_Chemical sub-corpus had the largest number of lines (2 972 895), almost six times that of the HUNER_Disease sub-corpus (559 063). The test sets had approximately half the number of lines of the corresponding training sets (ratios from 0.44 to 0.50) but the development sets were much smaller (development/training set ratios from 0.16 to 0.17). The number of annotated entities in the training sets ranged from 3 062 in the HUNER cell sub-corpus to 114 579 in the HUNER_chemical sub-corpus.

We next examined text similarity between the corresponding training and development sets and corresponding training and test sets using the scattertext tool which plots the frequency of words and bi-grams (Supplemental file 4) in two text collections in a scatterplot. Overall, the frequencies in HUNER training versus development sets and training versus test sets appeared to be relatively different. In particular, we observed clusters of terms that had high frequency in the development or test set but low frequency in the corresponding training set for the cell, species, gene/protein and disease class.

---

[32] https://github.com/Aitslab/EasyNER/blob/main/supplementary/experiment_scripts/preprocess_pubtatorformat.py

[33] https://github.com/Aitslab/EasyNER/blob/main/tutorials/Tutorial-benchmarking_entitylevel.md

**Training of BioBERT models for NER of genes/proteins, cells, species, diseases and chemicals on the HUNER corpora**

For model training, we combined the corresponding HUNER training and development sets to increase size and diversity of the training data

As language model, we chose BioBERT (v.1.1), which has shown very good performance when fine-tuned for different BioNLP tasks [17] but can be trained without excessive resources, in line with our ambition to make our research sustainable and easily reproducible. Fine-tuning was performed using the script from the BioBERT authors and their reported hyperparameters [17]. We used the cased BioBERT models (v1.1, PyTorch), as these perform slightly better according to their developers. Both BioBERT base and BioBERT large models were initially tested. However, the large models performed similarly to the base models (Table 1, data not shown) but required longer training and prediction times and were therefore not used further.

When evaluating the fine-tuned BioBERT base models on IOB2-token level, we obtained F1 scores between 0.64 and 0.88 for the five different entity classes (Table 1). Training corpus size was not clearly correlated with performance. For example, gene and species greatly differ in numbers of lines and entities, yet their F1 scores were almost the same. Nevertheless, the model trained on the smallest sub-corpus (BioBERT_HUNER_cell) had the lowest F1 score (0.64) suggesting that training data size might have been a limiting factor.

Table 1. IOB2-token-level evaluation of the fine-tuned BioBERT_HUNER base models. All _v1 models were trained on the HUNER train_dev sets with early stopping based on test set F1 score whereas all _v2 models were trained on the HUNER train sets only with early stopping based on dev set F1 score. Evaluation scores for the HUNER models[34] [29] are listed for comparison but these values represent the macro average of the scores (calculated by averaging the scores from each individual test set in the sub-corpus) and are thus not fully comparable with our scores which were calculated for the pooled sub-corpus test set. Prec = Precision, Rec = Recall, F1 = F1 score.

| Model | HUNER train | HUNER dev | HUNER test | | | HUNER test (macro average) | | |
|---|---|---|---|---|---|---|---|---|
| | F1 | F1 | Prec | Rec | F1 | Prec | Rec | F1 |
| BioBERT_ HUNER_cell_v1 | 1.00 | 1.00 | 0.65 | 0.68 | 0.66 | | | |
| BioBERT_HUNER _cell_v2 | 0.99 | 0.71 | 0.63 | 0.63 | 0.63 | | | |
| BioBERT_ HUNER_chemical_ v1 | 1.00 | 1.00 | 0.87 | 0.88 | 0.88 | | | |
| BioBERT_HUNER _chemical_v2 | 1.00 | 0.88 | 0.88 | 0.88 | 0.88 | | | |
| BioBERT_HUNER _disease_v1 | 1.00 | 1.00 | 0.85 | 0.84 | 0.85 | | | |
| BioBERT_HUNER 2_disease_v2 | 1.00 | 0.84 | 0.83 | 0.84 | 0.83 | | | |

[34]Retrieved from https://github.com/hu-ner/huner/blob/master/README.md on January 10, 2023.

| BioBERT_HUNER _gene_v1 | 0.99 | 1.00 | 0.76 | 0.78 | 0.77 | | | |
|---|---|---|---|---|---|---|---|---|
| BioBERT_HUNER _gene_v2 | 0.99 | 0.77 | 0.75 | 0.79 | 0.77 | | | |
| BioBERT_HUNER _species_v1 | 0.98 | 0.98 | 0.79 | 0.76 | 0.77 | | | |
| BioBERT_HUNER _species_v2 | 1.00 | 0.82 | 0.80 | 0.72 | 0.76 | | | |
| HUNER_cell | | | | | | 0.7 | 0.65 | 0.68 |
| HUNER_chemical | | | | | | 0.83 | 0.8 | 0.82 |
| HUNER_disease | | | | | | 0.75 | 0.78 | 0.76 |
| HUNER_gene | | | | | | 0.72 | 0.76 | 0.74 |
| HUNER_species | | | | | | 0.78 | 0.75 | 0.73 |

As expected, the BioBERT_HUNER_cell models recognized both generic cell terms and cell line names. For this model, partial matches which reflected differences in annotation practices rather than true errors were common (e.g. for "MG-63 cells" and "LNCaP cells" the ground truth did not include the word "cells" but the model prediction did). In addition, many instances counted as false positives were general terms referring to cells that had not been annotated in the ground truth data (e.g. tumor-derived cell lines, GFP-expressing parental cell line, fibroblast cell line).

The BioBERT_HUNER_chemical models recognized both therapeutic drugs and other chemicals and the BioBERT_HUNER_gene models recognized both genes and proteins as well as gene/protein family names (e.g. MAPK, ERK) (Figure 4A). The BioBERT_HUNER_disease models recognized disease names and terms closely related to diseases such as "tumor". The BioBERT_HUNER_species models recognized Linnean and common names.

As training data annotations were not designed for NER of nested entities, such entities were truncated as expected. For example in the sentence "(6E,13E)-18-bromo-12-butyl-11-chloro-4,8-diethyl-5-hydroxy-15-methoxytricosa-6,13-dien-19-yne-3,9-dione, 3-carboxy-3-hydroxypentanedioic and lactic acid are three chemicals." the second entity detected was "3-carboxy-3-hydroxypentanedioic" (whereas the fully correct entity would be "3-carboxy-3-hydroxypentanedioic acid").

## Generalization of HUNER-trained BioBERT models

Next, we evaluated the BioBERT_HUNER_disease, _species and _gene models on IOB2-token-level on two fully independent test sets, the Simplified Lund COVID-19 corpus (Table 2) and the CRAFT corpus (Table 3), to determine their ability to generalize. For the Simplified Lund COVID-19 corpus, we also evaluated the publicly available HunFlair [32] and ScispaCy models [37] for comparison (Table 2).

The BioBERT_HUNER_gene model performed relatively well, with an F1 score close to the one seen on the HUNER_gene test set (0.69 vs 0.77). In contrast, the BioBERT_HUNER_disease and BioBERT_HUNER_species models had much lower F1 scores on the simplified Lund COVID-19 corpus than on the respective HUNER test sets. Many of the false positive disease terms causing the low precision of the BioBERT_HUNER_disease models referred to symptoms (e.g. cough, fever), which were annotated as disease entities in the HUNER subcorpus used for training but not in the Simplified Lund COVID-19 corpus. Many of the false negative species terms causing low recall of the BioBERT_HUNER_species models referred to (corona)virus (e.g. coronavirus, 2019-nCoV, virus) or humans (e.g. human, patient). The BioBERT models outperformed the ScispaCy models for all entity classes, with the difference being especially large for species detection. HunFlair was evenly matched with our BioBERT models for the "Diseases" and "Species" entities but performed slightly worse for "Genes/Proteins".

Table 2. IOB2-token-level evaluation of the fine-tuned BioBERT_HUNER models on the Simplified Lund COVID-19 corpus. ScispaCy [37] and HunFlair [32] models were evaluated for comparison. Corpora that had been used for fine-tuning the ScispaCy models are indicated in the model name suffix.

| Model | Evaluated entity | Lund COVID-19 Precision | Lund COVID-19 Recall | Lund COVID-19 F1 score |
|---|---|---|---|---|
| BioBERT _HUNER_disease_v1 | Diseases | 0.29 | 0.55 | **0.38** |
| BioBERT _HUNER_disease_v2 | Diseases | 0.25 | 0.58 | 0.35 |
| ScispaCy en_ner_bc5cdr_md | Diseases | 0.20 | 0.50 | 0.29 |
| HunFlair | Diseases | | | **0.38** |
| BioBERT _HUNER_gene_v1 | Genes/Proteins | 0.81 | 0.76 | **0.79** |
| BioBERT _HUNER_gene_v2 | Genes/Proteins | 0.71 | 0.71 | 0.71 |
| ScispaCy en_ner_bionlp13cg_md | Genes/Proteins | 0.13 | 0.65 | 0.22 |
| ScispaCy en_ner_jnlpba_md | Genes/Proteins | 0.23 | 0.65 | 0.34 |
| ScispaCy en_ner_craft_md | Genes/Proteins | 0.04 | 0.29 | 0.07 |
| HunFlair | Genes/Proteins | | | 0.71 |
| BioBERT _HUNER_species_v1 | Species | 0.57 | 0.23 | **0.33** |
| BioBERT _HUNER_species_v2 | Species | 0.53 | 0.14 | 0.22 |
| ScispaCy en_ner_bionlp13cg_md | Species | 0.39 | 0.28 | **0.33** |
| ScispaCy en_ner_craft_md | Species | 0.23 | 0.24 | 0.23 |
| HunFlair | Species | | | 0.21 |

On the CRAFT corpus (Table 3), BioBERT_HUNER_v1 models showed reduced F1 scores for chemical entities (0.58 vs 0.88) compared to the HUNER_chemical test set. In contrast, the F1 scores for Gene/Protein and Species entities were almost identical on the two datasets. HunFlair performance was superior to our BioBERT_HUNER models on the CRAFT corpus.

Table 3. F1 scores from the IOB2-token-level evaluation of the three fine-tuned BioBERT_HUNER_v1 models on the CRAFT corpus. HunFlair [32] was evaluated for comparison. HunFlair results for the CRAFT corpus were better than those reported in the original paper due to differences in the evaluation procedure. Note that the entire CRAFT corpus, not just a test subset, was used for the evaluation.

| | CRAFT | | |
|---|---|---|---|
| | **Chemical** | **Gene/Protein** | **Species** |
| HunFlair | 0.85 | 0.89 | 0.96 |
| BioBERT_HUNER_v1 | 0.58 | 0.76 | 0.78 |

Lastly, we explored further whether fine-tuning BioBERT on the HUNER sub-corpora improves generalization compared to fine-tuning on an individual corpus. For this, we trained a BioBERT_base_cased_v1.1 model on the BC5CDR_disease corpus (Table 4). Training on the BC5CDR_disease corpus was performed the same way as for the V1 models by training on train_dev sets and evaluating on test sets. On the BC5CDR_disease test set, the BioBERT_BC5CDR_disease_v1 model had an IOB2-token-level F1 score similar to the BioBERT_HUNER_disease_v1 model and previously reported BioBERT models trained on the BC5CDR disease corpus. This F1 score was only slightly lower than that reported for BioMegatron, which has a different architecture. The precision of our BioBERT_BC5CDR_disease_v1 model was slightly higher than that of all these other models. For disease entity recognition on the Simplified Lund COVID-19 corpus and on the HUNER test set, the BioBERT_BC5CDR_disease model performed significantly worse than the BioBERT_HUNER_disease model. This suggests that training on the larger HUNER corpus collection indeed improved generalization.

Table 4. Generalization of BioBERT model trained on HUNER corpus collection vs single corpus. BioBERT models trained in the same manner on either the BC5CDR_disease corpus or the HUNER_disease sub-corpus (which includes the BC5CDR_disease corpus) were compared in their token-level performance. Published performance results for BioBERT models which were trained on the same single corpus are shown to confirm that our BioBERT_BC5CDRdiseases model was trained appropriately. For comparison, reported results from the state-of-the-art Megatron model trained on the BC5CDR corpus are also included. Note that the evaluation procedure for the published models differed slightly.

| | **HUNER_disease Test set** | | | **BC5CDR_disease Test set** | | | **Simplified Lund COVID-19 (disease entities)** | | |
|---|---|---|---|---|---|---|---|---|---|
| **Model** | **Prec** | **Rec** | **F1** | **Prec** | **Rec** | **F1** | **Prec** | **Rec** | **F1** |
| BioBERT_ HUNER_disease_v1 | 0.85 | 0.84 | 0.85 | 0.86 | 0.85 | 0.86 | 0.29 | 0.55 | 0.38 |
| BioBERT_ BC5CDR_disease_v1 | 0.79 | 0.70 | 0.75 | 0.87 | 0.86 | 0.86 | 0.25 | 0.48 | 0.33 |
| BioBERT_BC5CDR_disease_ Kuhnel [53] | | | | 0.82 | 0.85 | 0.83 | | | |
| BioBERT_BC5CDR_diseases_ Lee [17] | | | | 0.86 | 0.88 | 0.87 | | | |
| BioMegatron [15] | | | | 0.86 | 0.91 | 0.89 | | | |

### Development of the EasyNER end-to-end NER pipeline

Next, an end-to-end pipeline was designed to automatically process medical research articles from different sources with the five BioBERT NER models (Figure 1). In addition, a dictionary-based NER module and three COVID-19-related dictionaries we had developed previously were included [30, 31]. To add additional types of entities, we made it possible for users to incorporate their own BioBERT/BERT-like models or dictionaries. For flexibility and ease of use, we also included data loaders for a variety of inputs, including the entire PubMed database (which contains over 37 million abstracts) or a subset of it, a list of PubMed IDs (which can for example be obtained by exporting a search result), the CORD-19 metadata file (which contains over 750 000 COVID-19-related abstracts [3]) or a file with plain text. The pipeline consists of several processing modules that are run in sequence but can also be used individually. The modules reformat/download the desired text, split it into sentences and predict and quantify named entities.

The final output consists of a ranked list of extracted entities and a graph showing the top 50 entities, which provides a clear overview of the results. In addition, the pipeline generates JSON files with all text and detected entities (including their exact position) that can be used in downstream applications. For cases where the user wants to run more than one NER model, an optional merging module is included, which combines and compares the individual output files. We also included an accompanying free-standing script which allows the user to quickly inspect results for a specific entity.

**Benchmarking NER performance of EasyNER against other BioNLP tools**

To evaluate EasyNER, we compared it to other BioNLP tools that detect the same entities, including scispaCy [37], HunFlair2 [49], PubTator Central [50], PubTator 3.0 [27], BENT [51] and BERN2 [28].

As we had already assessed NER performance of the individual BioBERT models in EasyNER on IOB2-token level, we first compared entity-level performance. The evaluation was performed on several fully independent corpora ("cross-corpus evaluation") to obtain a better sense of the generalization ability and avoid the bias that is observed when test sets come from the same corpus as the training data ("in-corpus evaluation"): the tmVar 3.0 [39, 40], Bio-ID [41] and BioRED (test set only) [52] corpora and a simplified version of MedMentions [38], which we created by merging classes to match those in EasyNER. Note that BioRED has overlap with tmVar 3.0 [49].

EasyNER predictions were made with each of the _v1 models and predictions from the other tools were obtained from the HunFlair2 GitHub repository. In addition, new predictions were made with HunFlair2 (referred to as "HunFlair2 rerun"). PubTator 3.0 [27], an updated version of PubTator Central, was excluded from the NER benchmarking because the underlying model AIONER [54] was trained on most of the evaluation corpora. For HunFlair2, metrics for BioRED and tmVar 3.0 were excluded from benchmarking due to being "in-corpus".

First, we evaluated the entity-level performance for the "Cell" class (Figure 2A, B, Supplemental file 7). F1 scores varied greatly between corpora, which was not observed to this extent for other classes. On the Simplified MedMentions corpus, all tools had poor recall and consequently low F1 scores (from 0.04-0.16), with HunFlair2 and PubTator Central having the lowest scores. On the Bio-ID and tmVar 3.0 corpora, the performance of EasyNER, HunFlair/HunFlair2 re-run and PubTator Central was much better, whereas BENT and BERN2 still performed poorly. The highest F1 score within the "Cell" class was seen with EasyNER on the BioRED corpus, 0.77, which was only slightly below the in-corpus F1 score of the HunFlair2 rerun (0.85).

Next, we assessed the “Gene/Protein” class (Figure 2A, C, Supplemental file 7). EasyNER had the highest F1 score (0.64) for the Bio-ID corpus, with the other tools close behind. On the MedMentions corpus all tools except PubTator Central performed similarly well with F1 scores around 0.6. However, PubTator Central had the highest F1 score seen for the “Gene/Protein” class across all corpora on the tmVar 3.0 corpus (0.89), followed closely by BENT (0.83) and EasyNER (0.81). The second highest score overall for this class was for EasyNER on the BioRED corpus (0.85), again reaching a score only slightly below the in-corpus F1 score of the HunFlair2 rerun (0.95).

After this, we examined the performance on the “Chemical” class (Figure 2A, D, Supplemental file 7). On the Bio-ID corpus, all tools except PubTator Central had F1 scores of ~0.6. On the Simplified MedMentions corpus, HunFlair2/ HunFlair2 rerun had an F1 score in the same range, with EasyNER, BERN2 and BENT only slightly lower. Both PubTator Central and ScispaCy had lower scores on this corpus. As for the “Cell” class, the highest F1 score across all corpora was seen with EasyNER on the BioRED corpus (0.84).

The next evaluation was for the “Species” class (Figure 2A, E, Supplemental file 7). EasyNER F1 scores were slightly below those of the other tools for the Bio-ID corpus, with only ScispaCy even lower. Similarly, lower F1 scores for EasyNER were also seen for the Simplified MedMentions and tmVar 3.0 corpora. In all cases, the lower F1 scores of EasyNER were a consequence of low recall which ranged from 0.24 on the Simplified MedMentions corpus to 0.40 on the Bio-ID corpus whereas precision was much higher (0.71-0.82). The EasyNER F1 score was again highest on the BioRED corpus, surpassing all F1 scores of the other tools on the Bio-ID and Simplified MedMentions corpora, but staying below their score on the tmVar3.0 corpus.

Lastly, we evaluated the recognition of the “Disease” class (Figure 2A, F, Supplemental file 7). On the Simplified MedMentions corpus, HunFlair2, BENT, BERN2 and EasyNER F1 scores were similar, while PubTator Central and ScispaCy scores were lower. The EasyNER F1 score on the BioRED corpus (0.83) surpassed all other F1 scores even for this entity class and was again relatively close to the in-corpus F1 score of the HunFlair2 rerun (0.93).

In summary, the performance and ranking of the NER tools varied greatly depending on the corpus and entity class. EasyNER was competitive among the tools, performing best in the gene/protein class and on the BioRED corpus.

**Effect of Post-processing module on NER performance of EasyNER**

EasyNER contains a Post-processing module which merges adjacent tokens for entities containing hyphens and brackets and thereby avoids errors from partial recognition. To quantify the impact of this, we compared EasyNER performance with and without the Postprocessing module. Post-processing typically reduced the number of predicted entities by less than 1% and only led to improvements in F1 score of less than 0.005. It can thus be removed when processing speed is critical.

### EasyNER provides an excellent user experience for life scientists

To make BioNLP accessible to a wider audience, the tools need to be usable without extensive programming and NLP expertise, which most professionals in medicine and life science lack. They should also incorporate features that are of importance for this more general medical/life science end user group. We therefore made a qualitative evaluation of the EasyNER tool based on this perspective (Box 1).

**Box 1. EasyNER capabilities**

***Ease of use***

- runs on standard laptops with multiple operating systems
- integrated with PubMed and CORD-19
- in-built detection of common life science entities
- no prior programming or NLP expertise required
- extensive documentation with step-by-step tutorials
- in-built statistical analysis
- production of publication-ready graphs and ranked result tables

***Flexibility/customization***

- multiple types of input
- multiple types of output
- both neural network and dictionary-based NER
- can load user-generated dictionaries and models
- can process any text
- two tokenizers
- easy to add custom modules

***Control/transparency***

- user has full control over input data and NER method
- sentence-level traceability of results
- preservation of document metadata
- config file can be re-used and shared for reproducibility
- offline processing suitable for sensitive data
- disclosure of all model training materials and procedures
- all code, dictionaries and models available with open-source license
- step-by-step tutorials to repeat evaluations

***High performance NER***

- high-quality NER models
- no size limitations
- batching for very large text collections
- suitable for running in parallel on HPC clusters
- stable access due to local installation
- modular design
- production of annotated JSON files for further processing

**Deployment of the NER pipeline for autophagy-related information extraction from PubMed**

The EasyNER pipeline was tested on realistic text mining applications using our BioBERT_HUNER_v1 models. The first use case was information extraction from scientific abstracts related to autophagy. Two sets of autophagy-related abstracts were identified through searches on PubMed. The first dataset contained 1000 abstracts related to the central autophagy modulator mammalian target of rapamycin (mTOR) and its upstream regulator hamartin (TSC1). The second set contained 8333 abstracts related to the role of autophagy in cancer. Using the pipeline, we obtained downloaded the abstracts (Lund Autophagy-1 and Lund Autophagy-2 dataset, respectively) and performed NER with each of the five BioBERT models to detect cell, disease, chemical, species, and protein/gene entities.

*Protein/gene entities*

In the Lund Autophagy-1 dataset, mtor and tsc1, the abbreviated protein names used as search terms, were the most frequent entities detected by the BioBERT_HUNER_gene model (Figure 3A). In addition, several synonyms for these proteins were seen among the 50 most frequent entities, e.g. mammalian target of rapamycin and hamartin. Other frequent entities were abbreviated names of well-known genes/proteins or protein complexes that are in the same signaling pathway as mTOR and TSC1 such as mtorc1, tsc2, akt, rheb, pi3k, pten, ampk, s6k1. Full-length names of some autophagy regulators were also among the 50 most frequent entities as (e.g. tuberin, insulin) but not as many. Many of these frequently detected genes/proteins are part of the “mTOR signaling pathway” from the KEGG pathway database [55] (Figure 3B). We also detected some autophagy regulators not in the KEGG pathway (e.g. vegf, stat3, p53, tfe3, ghrelin, actin, c-myc, plk2).

In the autophagy/cancer-focused Lund Autophagy-2 dataset, mtor was also the most frequent protein/gene entity (Figure 4A). Several other frequent entities were also shared with the Lund Autophagy-1 dataset (e.g. akt, pi3k, mtorc1, ampk, p53). In addition, the 50 most common entities included autophagy receptors (e.g. p62/sqstm1) and parts of the autophagy-controlling atg conjugation system (e.g. lc3, atg5, ulk1, atg7). Some of the frequently found proteins/gene entities were also well-known oncogenes or tumor suppressors (e.g. akt, pi3k, p53).

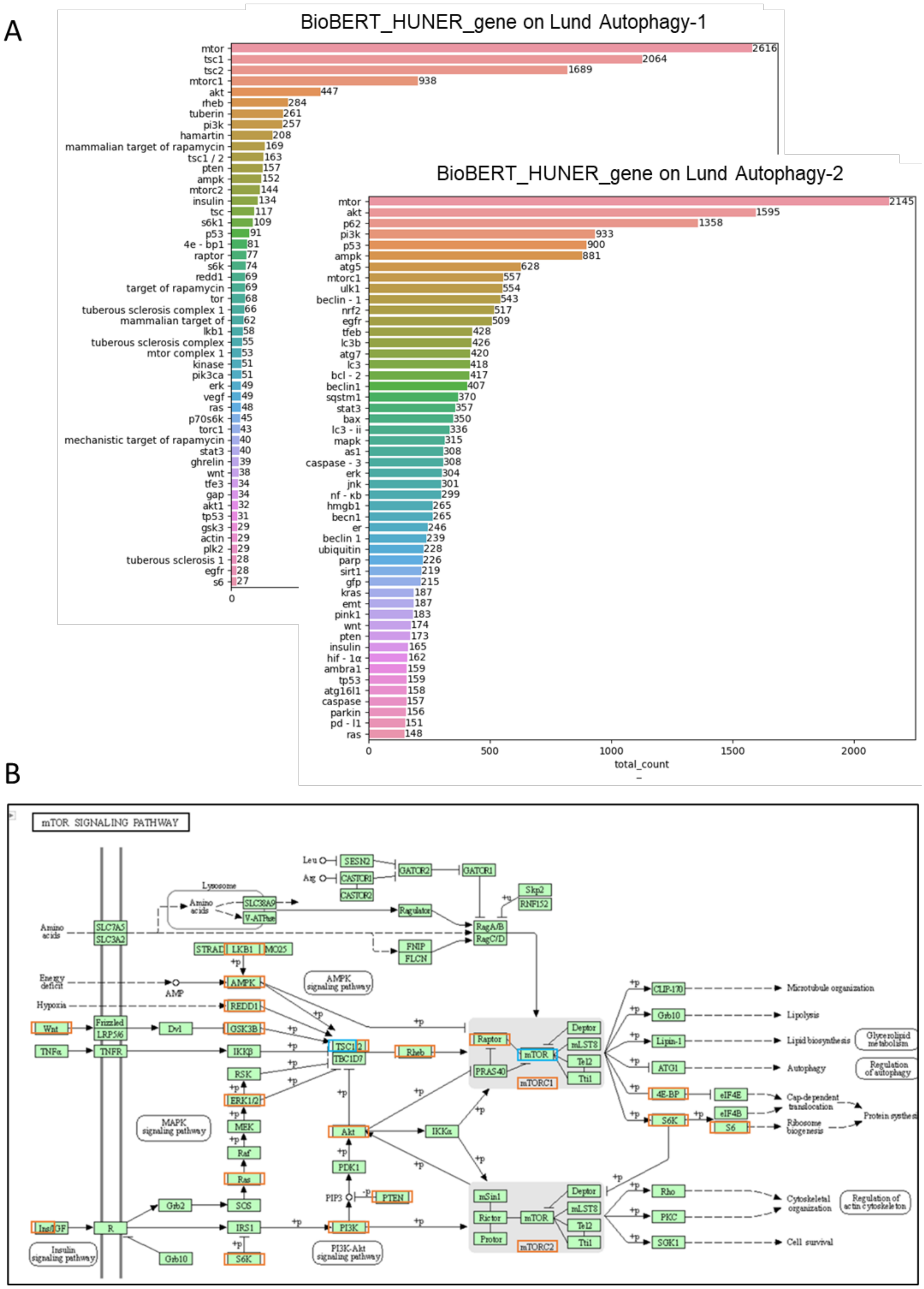

Figure 3. Frequent gene/protein entities in the autophagy-related datasets. A) 50 most frequent entities detected by the gene/protein model in the mTOR/TSC1-related Lund Autophagy-1 dataset and the autophagy/cancer-related Lund Autophagy-2 dataset. The model is a PyTorch BioBERT_cased_v1.1 model [17] fine-tuned on the HUNER gene sub-corpus (BioBERT_HUNER_gene). B) Overlap between mTOR signaling pathway and protein/gene entities detected in the Lund Autophagy-1 dataset. The mTOR signaling pathway was retrieved from KEGG database on 2023-01-09. The two proteins used as search terms to produce the dataset are highlighted in blue and other frequent entities in orange.

*Cell entities*

The BioBERT_HUNER_cell model falsely recognized the abbreviated protein/gene name tsc2 as the most frequent entity in the Lund Autophagy-1 dataset (Figure 4A). One of the other most common entities was the highly similar term "tsc2 -/- cells", which correctly indicated a type of cell (cell lacking the tsc2 gene). This prompted us to inspect the incorrect predictions more closely and we could see that when only "tsc2" was detected as entity it was typically part of a cell term that had not been identified in full, e.g. in the sentences "The augmented αB-crystallin was critical for the migration, invasion and apoptotic resistance of Tsc2-defective cells." or "This study shows that angiomyolipoma-derived human smooth muscle TSC2-/- cells express the apoptosis inhibitor protein survivin when exposed to IGF-1."

Other highly frequent terms represented true cell terms.

In the Lund Autophagy-2 dataset, many well-known cancer cell lines were detected frequently. As expected, spelling variants were picked up for several of them (e.g. a549/a549 cells, mcf-7/mcf7) (Figure 4A). However, two cancer type abbreviations, nsclc (= non-small cell lung cancer) and crc (= colorectal cancer) were also wrongly listed among the 50 most frequent cell entities. As for tsc2 in the Autophagy-1 dataset, many instances were longer incompletely recognized cell terms that included crc or nsclc.

*Chemical entities*

The BioBERT_HUNER_chemical model detected the mTOR inhibitors rapamycin and everolimus and the rapamycin brand name sirolimus as three of the five most frequent entities in the Lund Autophagy-1 dataset (Figure 4B). Autophagy-regulating metabolites that act through the mTOR signaling pathway were also frequently detected (e.g. glucose, amino acids).

In the Lund Autophagy-2 dataset, many chemicals belonging to one of three groups were found (Figure 4B): 1. Anti-cancer chemotherapy agents (e.g. cisplatin, dox/doxorubicin, sorafenib), 2. Autophagy-modulating drugs (e.g. rapamycin, chloroquine) and 3. Basic chemicals/metabolites (e.g. oxygen, glucose, iron, atp).

*Disease entities*

The BioBERT_HUNER_disease model found tsc as the most frequent entity in the Lund Autophagy-1 dataset (Figure 4C). tsc is an abbreviation for "tuberous sclerosis complex", a disease caused by mutations in TSC1 (the gene symbol used as search term). The full name and the synonym tuberous sclerosis were also detected with very high frequency. The other most common disease terms were tumor, tumors, cancer, epilepsy and seizures. As tumors and seizures are common in patients with tuberous sclerosis complex, these terms were also expected to rank highly.

In the Lund Autophagy-2 dataset (Figure 4C), most of the 15 most frequent disease entities were terms for cancers, as would be expected from the cancer-focused article selection for this dataset. The model was able to recognize both full names and common abbreviations (e.g. crc = colorectal cancer, hcc = hepatocellular carcinoma, nsclc = non-small cell lung cancer, gbm = glioblastoma multiforme).

*Species entities*

Finally, all top-ranked entities predicted by the BioBERT_HUNER_species model in the Lund Autophagy-1 dataset were indeed terms referring to species (Figure 4D). This included model organisms (e.g. mice, mouse, rat), terms referring to humans (e.g. patient, patients, children), species-describing adjectives (e.g. murine), and abbreviated virus names (e.g. hcv, hbv). Similar results were seen with the Lund Autophagy-2 dataset.

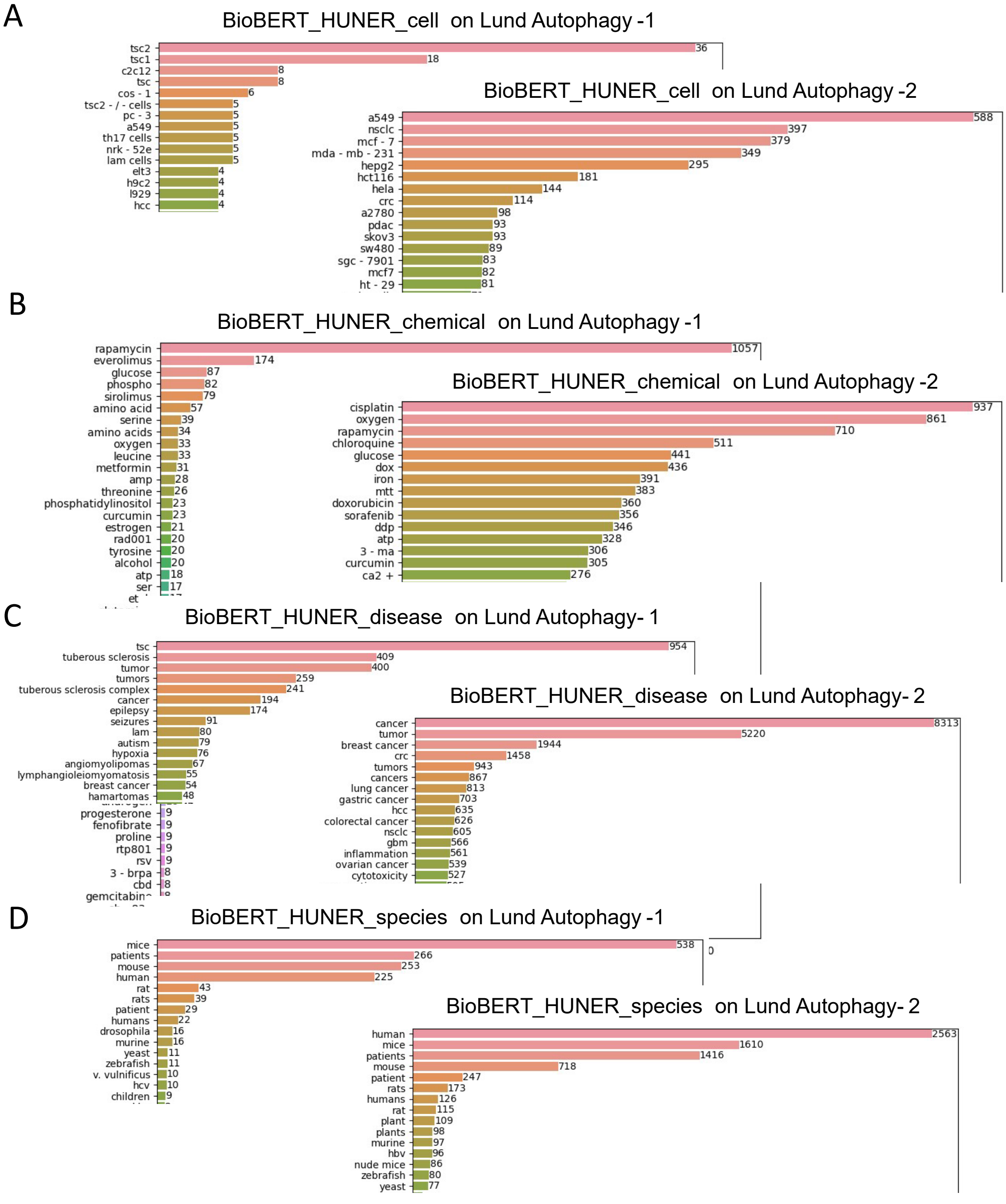

Figure 4. 15 most frequent entities detected by the (A) BioBERT_HUNER_cell model, (B) BioBERT_HUNER_chemical model, (C) BioBERT_HUNER_disease model and (D) the BioBERT_HUNER_species model in the autophagy-related datasets. The models are PyTorch BioBERT_cased_v1.1 models fine-tuned on the respective HUNER sub-corpus. The numbers on the bar plots indicate the number of times the detected entity occurs within the respective corpus.

#### Deployment of the NER pipeline for information extraction from CORD-19

As second use case, the pipeline was deployed on CORD-19, a collection of COVID-19-related articles (Figure 5) [3]. Titles and abstracts from the over 700 000 unique CORD-19 records were extracted from the metadata file. The protein/gene model correctly identified many relevant proteins/gene terms (igg, ace2, cytokine, il-6) but mistakenly included COVID-19 in this entity class. Similarly, the cell model misidentified many variants of the term covid-19 in addition to correctly detecting cell lines widely used for COVID-19 research (e.g. vero e6, a549, calu3). The BioBERT_HUNER_chemical model identified oxygen, alcohol and glucose as most commonly found hits. Other frequent entities were the therapeutic drugs, that had been explored as treatments, e.g. hydroxychloroquine, vitamin d, remdesivir and dexamethasone. The disease model identified several terms directly associated with COVID-19 (e.g. infection, coronavirus disease, pneumonia, sars-cov-2 infection, covid-19, acute respiratory syndrome) among the most common entities. Other most frequent entities were common diseases such as anxiety, cancer, depression, diabetes. The species model most frequently found terms describing humans, model organisms, the SARS-CoV2 virus and other viruses.

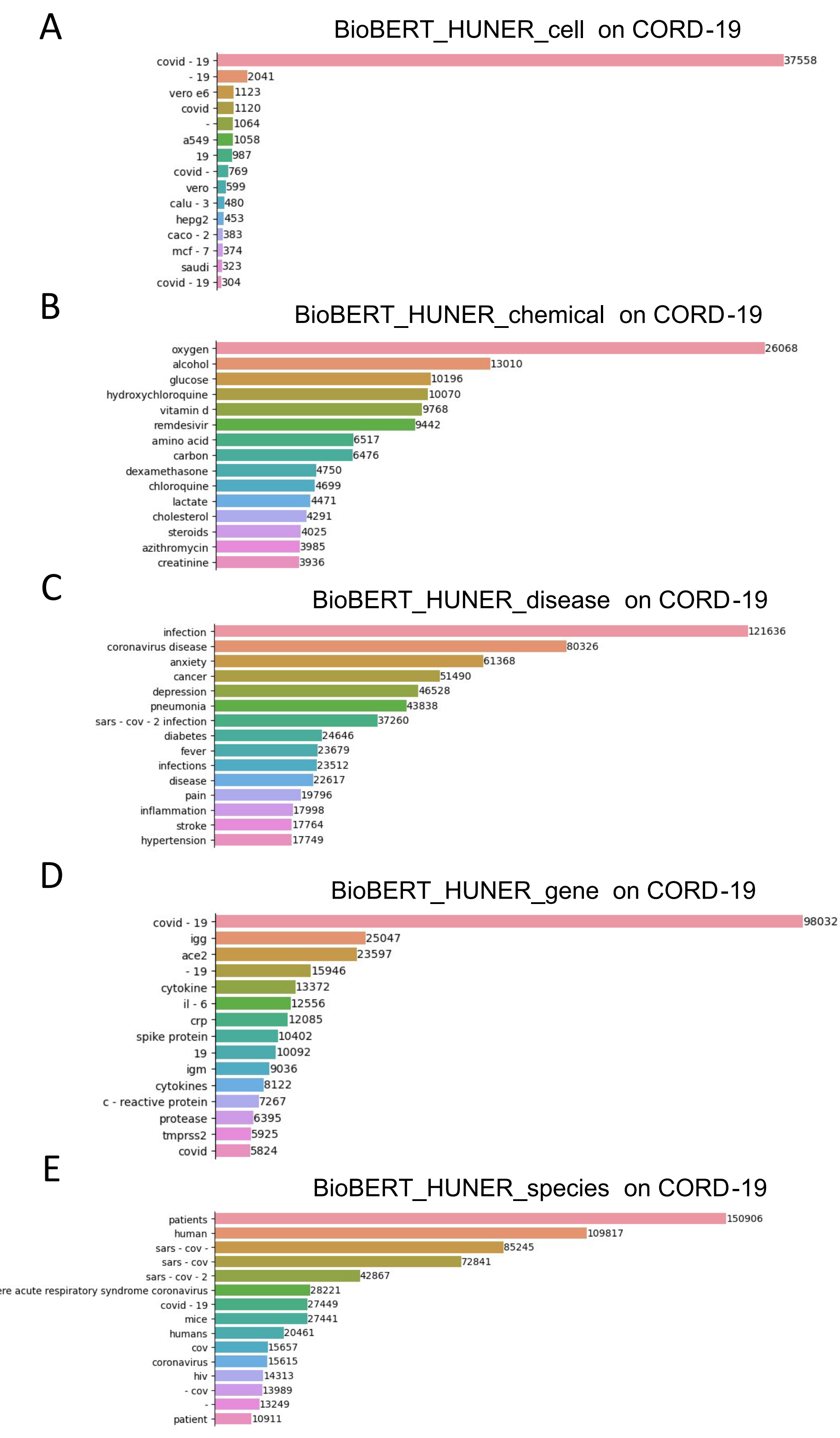

Figure 5. 15 most frequent entities detected by the A) BioBERT_HUNER_cell, B) BioBERT_HUNER_chemical, C) BioBERT_HUNER_disease D) BioBERT_HUNER_gene and E) BioBERT_HUNER_species models on the CORD-19 abstracts [3]. The models are PyTorch BioBERT_cased_v1.1 models fine-tuned on the respective HUNER sub-corpus. The numbers on the bar plots indicate the number of times the detected entity occurs within the respective corpus.

## Discussion

We developed an end-to-end NER pipeline for information extraction from medical and life science texts called EasyNER which incorporates both deep neural network- and dictionary-based approaches. It has in-built capabilities to connect to major collections of research articles (PubMed, CORD-19) and detect terms of broad interest and was designed to provide high performance NER with full control, transparency, flexibility and customization possibilities while maintaining ease of use for medical and life science professionals (Box 1).

The included models were based on BioBERT which has excellent performance on medical NER tasks [17]. As generalization is a major concern, we fine-tuned BioBERT models on the HUNER collection, which aggregates multiple corpora for each entity. Similarly to our work, the HUNER sub-corpora have recently been used to train HunFlair [32]. HunFlair performed better than our models on the CRAFT corpus but equally or worse for the different entity classes in the Simplified Lund COVID-19 corpus. However, as the HunFlair models were trained on the combined HUNER training and test sets, whereas our BioBERT models were trained on the combined training and development sets, a direct comparison was not possible.

While the Flair library is easy to use for NLP experts it is not targeted toward life scientists. The same is true for the ScispaCy models, another set of pre-trained medical NER models [37]. ScispaCy models were only trained on single corpora, however, and performed more poorly than our models on the Simplified Lund COVID-19 corpus. In contrast to our pipeline, HunFlair [32] and ScispaCy do not allow direct access of medical article collections such as PubMed and CORD-19 [3].

Our pipeline generates a ranked list and bar graph, providing an easy results overview and publication-ready files. For cases, where multiple entity types are detected in sequential runs with separate models, the pipeline includes a module ("Merge entities") that combines and compares the predictions.

Our pipeline has many applications that can support medical research. For example, it can give life scientists insight into proteins/genes reported to participate in a specific cell process or signaling pathway. Today, life scientists often rely on pathway databases but these are incomplete. By performing NER on research articles related to autophagy-regulators, we could thus detect regulators mentioned on the mTOR signaling map in the KEGG database but also several key regulators that were not in the map (e.g. p53, vegf, stat3, tfe3, ghrelin).

A second application of our pipeline is to quickly gain an overview of experimental tools. When using the pipeline, we could reveal commonly used cell lines, experimental drugs (e.g. mTOR inhibitors) and model organisms for autophagy and COVID-19-related research. Here again, the filtering of input articles allows for more nuanced insights.

Similarly, researchers can use our pipeline to identify drug candidates for a disease and assess the number of articles published on them. As expected, extensively studied drug candidates for COVID-19, e.g. hydroxychloroquine, remdesivir and dexamethasone, were among the top 50 most frequent entities detected by the chemical NER pipeline in the CORD-19 abstracts [3]. Alternatively, users could rapidly find proteins/genes mentioned in articles about a specific drug.

These are just a few use examples. For many of these information tasks, excellent bioinformatics databases do exist (e.g. protein-protein interaction or gene-disease databases), but these are typically incomplete as many rely on manual curation. They are also time-consuming to explore as they do not give the user the ability to target their search in the same manner as our NER pipeline. The NER pipeline is thus an excellent complement to existing databases, allowing the user to customize and speed up their search for information.

One limitation of the pipeline is that it does not perform named entity linking. Multiple spelling variants and synonyms were thus not merged. In many scenarios, there is a dominant spelling variant, however, and often variants can easily be identified and harmonized in post-processing (e.g. by removing hyphens). The conversion to lower case performed by the pipeline at least eliminated capitalization variants. Another limitation is that we, like most other NER tools, did not train our models to resolve nested entities. Consequently, predictions truncate the first entity in a nested expression. Many users will be able to recognize the entities despite the truncation, however.

A limitation of the two case studies is that the ground truth is unknown. Manual inspection revealed that a common case of false negatives was the failure to recognize all parts of a multi-word entity. In contrast, there were few false positives among the 50 most frequent terms and those that were observed (e.g. COVID-19 as cell line) could easily be filtered out.

## Conclusion

Our end-to-end NER pipeline can help medical researchers with various information extraction tasks without requiring specialist NLP knowledge. It contains BioBERT NER models that recognize terms for cells, chemicals/drugs, diseases, genes/proteins, and species and dictionaries that can help find COVID-19 or SARS-CoV2 synonyms, including virus variant names. The pipeline can also incorporate models and dictionaries provided by the user, leading to great flexibility.

## Supplemental files

Supplemental files are available at: https://github.com/Aitslab/EasyNER/tree/main/supplementary

**Supplemental file 1.** Size of the HUNER sub-corpora.

**Supplemental file 2.** Size of the Simplified Lund COVID-19 corpus (after merging of original entity classes).

**Supplemental file 3.** Zip file with Jupyter notebook/Scripts for environment setup, gold standard corpus acquisition and pre-processing, model training and evaluation.

**Supplemental file 4.** Zip file with scattertext html files.

**Supplemental file 5.** BioBERT training curves

**Supplemental file 6.** Zip folder with text files containing examples of wrongly identified or missed entities of the BioBERT models

**Supplemental file 7.** Results of entity-level benchmarking

**Supplemental file 8.** Comparison of EasyNER performance with and without postprocessing

**Supplemental file 8**. Lund Autophagy-1 text collection

**Supplemental file 9**. Lund Autophagy-2 text collection

**Supplemental file 10**. Zip folder with output files of pipeline from Autophagy-1 text collection for all models

## Acknowledgement

The computations and data handling were enabled by resources provided by the National Academic Infrastructure for Supercomputing in Sweden (NAISS) and the Swedish National Infrastructure for Computing (SNIC) at Lund University (LUNARC), Chalmers University of Technology (Alvis), and the National Supercomputer Centre at Linköping University (Berzelius, provided by the Knut and Alice Wallenberg foundation), partially funded by the Swedish Research Council through grant agreements no. 2022-06725 and no. 2018-05973.

This study was supported by the Swedish Research Council, the SciLifeLab/Knut and Alice Wallenberg COVID-19 national research program, the Wallenberg AI, Autonomous Systems and Software Program – Humanities and Society (WASP-HS) and Data-driven life science (DDLS) program, the Swedish Research Council for Sustainable Development (FORMAS), the Crafoord Foundation and the Lund University Sustainability Fund.

We also acknowledge the following research environments and networks which support our work: AI Lund, AIR Lund, Lund University Profile Area "Nature-based Future Solutions", Lund University Profile Area "Natural and Artificial Cognition", Lund University Profile Area "Proactive Ageing", LTH Profile Area "AI and Digitalization", LTH Profile Area "Engineering Health", Strategic Research Areas "BECC", "EpiHealth" and "eSSENCE", and PhenoTarget.

The pipeline figure was designed using images from flaticon.com from authors Darius Dan https://www.flaticon.com/authors/darius-dan), eukalyp

(https://www.flaticon.com/authors/eucalyp), Freepik and pojok d (https://www.flaticon.com/authors/pojok-d).

## CRediT author statement

Rafsan Ahmed: Methodology, Software, Validation, Formal analysis, Investigation, Data Curation, Writing - Original Draft, Visualization, Supervision

Petter Berntsson: Methodology, Software, Validation, Formal analysis, Investigation, Data Curation

Alexander Skafte: Methodology, Software, Validation, Formal analysis, Investigation, Data Curation

Salma Kazemi Rashed: Methodology, Software, Validation, Formal analysis, Investigation, Data Curation, Supervision

Marcus Klang: Methodology, Software, Resources, Data Curation, Supervision

Adam Barvesten: Software, Formal analysis, Investigation, Data Curation

Ola Olde: Software, Formal analysis, Investigation, Data Curation

William Lindholm: Software, Formal analysis, Investigation

Antton Lamarca Arrizabalaga: Software, Formal analysis, Investigation

Pierre Nugues: Methodology, Supervision

Sonja Aits: Conceptualization, Methodology, Software, Validation, Formal analysis, Investigation, Resources, Data Curation, Writing - Original Draft, Writing - Review & Editing, Visualization, Supervision, Project administration, Funding acquisition

## Competing Interests Statement

The authors have no competing interests in relation to this article.

## References

1. NCBI: Pubmed. https://pubmed.ncbi.nlm.nih.gov/. Accessed 22nd November 2022.
2. Sayers EW, Bolton EE, Brister JR, Canese K, Chan J, Comeau Donald C, et al. Database resources of the national center for biotechnology information. Nucleic Acids Research. 2022;50 D1:D20-D6. doi:10.1093/nar/gkab1112.
3. Wang LL, Lo K, Chandrasekhar Y, Reas R, Yang J, Burdick D, et al. CORD-19: The COVID-19 Open Research Dataset. ArXiv. 2020; doi:https://doi.org/10.48550/arXiv.2004.10706.
4. Devlin J, Chang M-W, Lee K and Toutanova K. BERT: Pre-training of Deep Bidirectional Transformers for Language Understanding. In: Minneapolis, Minnesota, June 2019, pp.4171-86. Association for Computational Linguistics.
5. Chang L, Zhang RH, Lv J, Zhou WG and Bai YL. A review of biomedical named entity recognition. J Comput Methods Sci. 2022;22 3:893-900. doi:10.3233/Jcm-225952.
6. Perera N, Dehmer M and Emmert-Streib F. Named Entity Recognition and Relation Detection for Biomedical Information Extraction. Front Cell Dev Biol. 2020;8:673. doi:10.3389/fcell.2020.00673.
7. Cook HV and Jensen LJ. A Guide to Dictionary-Based Text Mining. Methods Mol Biol. 2019;1939:73-89. doi:10.1007/978-1-4939-9089-4_5.
8. Mikolov T, Chen K, Corrado G and Dean J. Efficient Estimation of Word Representations in Vector Space. arXiv. 2013:arXiv:1301.3781. doi:https://doi.org/10.48550/arXiv.1301.3781.
9. Pennington J, Socher R and Manning C. Glove: Global Vectors for Word Representation. *Proceedings of the 2014 Conference on Empirical Methods in Natural Language Processing (EMNLP)*. 2014, p. 1532-43.
10. Garneau N, Leboeuf J-S and Lamontagne L. Predicting and interpreting embeddings for out of vocabulary words in downstream tasks. *Proceedings of the 2018 EMNLP Workshop BlackboxNLP: Analyzing and Interpreting Neural Networks for NLP*. 2018, p. 331-3.
11. Chelba C, Mikolov T, Schuster M, Ge Q, Brants T, Koehn P, et al. One billion word benchmark for measuring progress in statistical language modeling. *Interspeech 2014*. 2014, p. 2635-9.
12. Radford A, Narasimhan K, Salimans T and Sutskever I. Improving language understanding by generative pre-training. 2018.
13. Ross Taylor MK, Guillem Cucurull, Thomas Scialom, Anthony Hartshorn, Elvis Saravia, Andrew Poulton, Viktor Kerkez and Robert Stojnic. GALACTICA: A Large Language Model for Science. arXiv. 2022:arXiv:2211.09085. doi:https://doi.org/10.48550/arXiv.2211.09085.
14. Vaswani A, Shazeer N, Parmar N, Uszkoreit J, Jones L, Gomez AN, et al. Attention is All you Need. arXiv. 2017:arXiv:1706.03762. doi:https://doi.org/10.48550/arXiv.1706.03762.
15. Shin H-C, Zhang Y, Bakhturina E, Puri R, Patwary M, Shoeybi M, et al. BioMegatron: Larger Biomedical Domain Language Model. 2020:arXiv:2010.06060.
16. Ruder S, Peters ME, Swayamdipta S and Wolf T. Transfer Learning in Natural Language Processing. Minneapolis, Minnesota: Association for Computational Linguistics; 2019.
17. Lee J, Yoon W, Kim S, Kim D, Kim S, So CH, et al. BioBERT: a pre-trained biomedical language representation model for biomedical text mining. Bioinformatics. 2020;36 4:1234-40. doi:10.1093/bioinformatics/btz682.
18. Alsentzer E, Murphy J, Boag W, Weng W-H, Jindi D, Naumann T, et al. Publicly Available Clinical BERT embeddings. *Proceedings of the 2nd Clinical Natural Language Processing Workshop*. 2019, p. 72-8.
19. Peng Y, Yan S and Lu Z. Transfer Learning in Biomedical Natural Language Processing: An Evaluation of BERT and ELMo on Ten Benchmarking Datasets. 2019:arXiv:1906.05474. doi:10.48550/arXiv.1906.05474.

20. Gu Y, Tinn R, Cheng H, Lucas M, Usuyama N, Liu X, et al. Domain-Specific Language Model Pretraining for Biomedical Natural Language Processing. 2020:arXiv:2007.15779. doi:10.48550/arXiv.2007.15779.
21. Haley C. This is a BERT. Now there are several of them. Can they generalize to novel words? *Proceedings of the Third BlackboxNLP Workshop on Analyzing and Interpreting Neural Networks for NLP*. 2020, p. 333-41.
22. Montani I, Honnibal M, Boyd A, Landeghem SV and Peters H. explosion/spaCy: v3.7.2: Fixes for APIs and requirements (v3.7.2). Zenodo, 2023.
23. Akbik A, Bergmann T, Blythe D, Rasul K, Schweter S and Vollgraf R. FLAIR: An Easy-to-Use Framework for State-of-the-Art NLP. In: Minneapolis, Minnesota, June 2019, pp.54-9. Association for Computational Linguistics.
24. Wolf T, Debut L, Sanh V, Chaumond J, Delangue C, Moi A, et al. HuggingFace's Transformers: State-of-the-art Natural Language Processing. 2019. doi:10.48550/arXiv.1910.03771.
25. Szklarczyk D, Gable AL, Lyon D, Junge A, Wyder S, Huerta-Cepas J, et al. STRING v11: protein-protein association networks with increased coverage, supporting functional discovery in genome-wide experimental datasets. Nucleic Acids Res. 2019;47 D1:D607-D13. doi:10.1093/nar/gky1131.
26. Ferguson C, Araujo D, Faulk L, Gou Y, Hamelers A, Huang Z, et al. Europe PMC in 2020. Nucleic Acids Res. 2021;49 D1:D1507-D14. doi:10.1093/nar/gkaa994.
27. Wei CH, Allot A, Lai PT, Leaman R, Tian S, Luo L, et al. PubTator 3.0: an AI-powered literature resource for unlocking biomedical knowledge. Nucleic Acids Res. 2024;52 W1:W540-W6. doi:10.1093/nar/gkae235.
28. Sung M, Jeong M, Choi Y, Kim D, Lee J, Kang J, et al. BERN2: an advanced neural biomedical named entity recognition and normalization tool. Bioinformatics. 2022;38 20:4837-9. doi:10.1093/bioinformatics/btac598.
29. Weber L, Munchmeyer J, Rocktaschel T, Habibi M and Leser U. HUNER: improving biomedical NER with pretraining. Bioinformatics. 2020;36 1:295-302. doi:10.1093/bioinformatics/btz528.
30. Kazemi Rashed S, Ahmed R, Frid J and Aits S. Files and code for English dictionaries, gold and silver standard corpora for biomedical natural language processing related to SARS-CoV-2 and COVID-19. zenodo. 2022.
31. Kazemi Rashed S, Ahmed R, Frid J and Aits S. English dictionaries, gold and silver standard corpora for biomedical natural language processing related to SARS-CoV-2 and COVID-19. 2020:arXiv:2003.09865. doi: https://doi.org/10.48550/arXiv.2003.09865.
32. Weber L, Sanger M, Munchmeyer J, Habibi M, Leser U and Akbik A. HunFlair: An Easy-to-Use Tool for State-of-the-Art Biomedical Named Entity Recognition. Bioinformatics. 2021;37 17:2792-4. doi:10.1093/bioinformatics/btab042.
33. Tjong Kim Sang EF. Introduction to the CoNLL-2002 Shared Task: Language-Independent Named Entity Recognition. In: 2002.
34. Furlong LI, Dach H, Hofmann-Apitius M and Sanz F. OSIRISv1.2: A named entity recognition system for sequence variants of genes in biomedical literature. BMC bioinformatics. 2008;9 1 doi:10.1186/1471-2105-9-84.
35. Li J, Sun Y, Johnson RJ, Sciaky D, Wei C-H, Leaman R, et al. BioCreative V CDR task corpus: a resource for chemical disease relation extraction. Database. 2016;2016 doi:10.1093/database/baw068.
36. Cohen KB, Verspoor K, Fort K, Funk C, Bada M, Palmer M, et al. The Colorado Richly Annotated Full Text (CRAFT) Corpus: Multi-Model Annotation in the Biomedical Domain. Handbook of Linguistic Annotation. 2017. p. 1379-94.
37. Neumann M, King D, Beltagy I and Ammar W. ScispaCy: Fast and Robust Models for Biomedical Natural Language Processing. In: Florence, Italy, aug 2019, pp.319-27. Association for Computational Linguistics.

38. Mohan S and Li D. MedMentions: A Large Biomedical Corpus Annotated with UMLS Concepts. 2019. doi:10.48550/arXiv.1902.09476.
39. Wei CH, Allot A, Riehle K, Milosavljevic A and Lu Z. tmVar 3.0: an improved variant concept recognition and normalization tool. Bioinformatics. 2022;38 18:4449-51. doi:10.1093/bioinformatics/btac537.
40. Wei CH, Harris BR, Kao HY and Lu Z. tmVar: a text mining approach for extracting sequence variants in biomedical literature. Bioinformatics. 2013;29 11:1433-9. doi:10.1093/bioinformatics/btt156.
41. Arighi C, Hirschman L, Lemberger T, Bayer S, Liechti R, Comeau D, et al. Bio-ID Track Overview. In: *BioCreative VI Challenge Evaluation Workshop* 2017, pp.14-9.
42. Kessler JS. Scattertext: a Browser-Based Tool for Visualizing how Corpora Differ. 2017:arXiv:1703.00565. doi:10.48550/arXiv.1703.00565.
43. Goyal P, Dollár P, Girshick R, Noordhuis P, Wesolowski L, Kyrola A, et al. Accurate, Large Minibatch SGD: Training ImageNet in 1 Hour. 2017:arXiv:1706.02677. doi:10.48550/arXiv.1706.02677.
44. Hiroki N. seqeval : A Python framework for sequence labeling evaluation. 2018.
45. Achakulvisut T, Acuna D and Kording K. Pubmed Parser: A Python Parser for PubMed Open-Access XML Subset and MEDLINE XML Dataset XML Dataset. Journal of Open Source Software. 2020;5 46 doi:10.21105/joss.01979.
46. Steven Bird EK, and Edward Loper. Natural Language Processing with Python. O'Reilly Media Inc.; 2009.
47. Honnibal M, Montani I, Van Landeghem S and Boyd A. spaCy: Industrial-strength Natural Language Processing in Python. 2020; doi:10.5281/zenodo.1212303.
48. Wu Y, Schuster M, Chen Z, Le QV, Norouzi M, Macherey W, et al. Google's Neural Machine Translation System: Bridging the Gap between Human and Machine Translation. 2016:arXiv:1609.08144. doi:10.48550/arXiv.1609.08144.
49. Sänger M, Garda S, Wang XD, Weber-Genzel L, Droop P, Fuchs B, et al. HunFlair2 in a cross-corpus evaluation of biomedical named entity recognition and normalization tools. 2024. doi:10.48550/arXiv.2402.12372.
50. Wei CH, Allot A, Leaman R and Lu Z. PubTator central: automated concept annotation for biomedical full text articles. Nucleic Acids Res. 2019;47 W1:W587-W93. doi:10.1093/nar/gkz389.
51. Pedro R and Francisco MC. NILINKER: Attention-based approach to NIL Entity Linking. Journal of Biomedical Informatics. 2022;132:104137. doi:https://doi.org/10.1016/j.jbi.2022.104137.
52. Luo L, Lai PT, Wei CH, Arighi CN and Lu Z. BioRED: a rich biomedical relation extraction dataset. Brief Bioinform. 2022;23 5 doi:10.1093/bib/bbac282.
53. Kuhnel L and Fluck J. We are not ready yet: limitations of state-of-the-art disease named entity recognizers. J Biomed Semantics. 2022;13 1:26. doi:10.1186/s13326-022-00280-6.
54. Luo L, Wei CH, Lai PT, Leaman R, Chen Q and Lu Z. AIONER: all-in-one scheme-based biomedical named entity recognition using deep learning. Bioinformatics. 2023;39 5 doi:10.1093/bioinformatics/btad310.
55. Kanehisa M, Furumichi M, Tanabe M, Sato Y and Morishima K. KEGG: new perspectives on genomes, pathways, diseases and drugs. Nucleic Acids Res. 2017;45 D1:D353-d61. doi:10.1093/nar/gkw1092.